\documentstyle[psfig]{mn}

\begin{document}
\title[Properties of the double $\beta$ model for intracluster gas]
{Properties of the double $\beta$ model for intracluster gas}  
\author[Xue and Wu]{Yan-Jie Xue and Xiang-Ping Wu\\
%\footnotesize{
Beijing Astronomical Observatory and National Astronomical Observatories,
Chinese Academy of Sciences, Beijing 100012, China\\
}

\date{Accepted 2000 June 8. in original form 1999 September 29}
\maketitle
\begin{abstract}
We present an extensive study of the double $\beta$ 
model for the X-ray surface brightness profiles of clusters, 
and derive analytically the gas density and total masses of clusters 
under the hydrostatic equilibrium hypothesis.
It is shown that the employment of the double $\beta$ model instead 
of the conventional single $\beta$ model can significantly improve 
the goodness of fit to the observed X-ray surface brightness profiles 
of clusters, which will in turn lead to a better determination of
the gas and total mass distributions in clusters. 
In particular, the observationally fitted $\beta$ parameter for 
the extended component in a double $\beta$ model 
may become larger. This opens a new possibility of  
resolving the long-standing $\beta$ discrepancy for clusters.
Using an ensemble of 33 ROSAT PSPC observed
clusters drawn from the Mohr, Mathiesen \& Evrard (1999) sample,
we find that the asymptotic value of $\beta_{fit}$ is $0.83\pm0.33$ 
at large radii, consistent with both the average spectroscopic parameter 
$\beta_{spec}=0.78\pm0.37$ and the result given by  numerical simulations. 
\end{abstract}


\vskip -3in

\section{Introduction}

Since the pioneering work of Cavaliere \& Fusco-Femiano (1976),  
the $\beta$ model has been widely adopted in the fitting of the
X-ray surface brightness profiles of galaxy clusters.
In particular, despite the difference in the core radius
there is a striking similarity between 
the X-ray surface brightness distribution predicted by 
the universal density profile as the underlying gravitational 
potential of a cluster and the
conventional $\beta$ model (Makino, Sasaki \& Suto 1998).
Yet,  it has been realized for many years that 
a single $\beta$ model is inaccurate and also not self-consistent. 
First, it fails to represent the central excess emission associated with
the cooling flows seen in many clusters. Second, the X-ray luminosity is 
divergent for $\beta<0.5$ so that an arbitrary cutoff radius should
be introduced (e.g. Henry \& Henriksen 1986). 
Third, an increasing temperature with cluster radius is
required for $\beta<2/3$ if the asymptotic baryon fraction of cluster
at large radii should reach a universal value defined by the Big Bang
Nucleosynthesis (Wu \& Xue 2000). Finally, the well-known $\beta$
discrepancy (e.g. Bahcall \& Lubin 1994) 
may also arise from the employment of a single $\beta$ model. 
Namely,  the slope of gas radial profile described
by a single $\beta$ model is systematically smaller than that
required by the equipartition between 
specific kinetic energy in galaxies and that in gas.

Regardless of its simplification, the second $\beta$ model has been
formally adopted by a number of authors in recent years to represent 
the excess X-ray emission 
in the central cores of  cooling flow clusters (e.g. Ikebe et al. 1996; 1999; 
Xu et al. 1998; Mohr, Mathiesen \& Evrard 1999, MME hereafter).  
An immediate consequence of fitting the X-ray surface brightness profile with
a double $\beta$ model instead of a single $\beta$ model 
is that the resultant $\beta$ parameters 
in the two components may both become larger. 
In particular, it seems  that the $\beta$
parameter for the extended component 
can often be greater than $2/3$. Therefore,
there is a possibility that a double $\beta$ model may allow one to
resolve all the above puzzles associated with a single $\beta$ model.
Moreover, in the conventional single $\beta$ model fit,
some of the observed data points in the central regions of 
the cooling flow clusters should be omitted 
in order to obtain an acceptable fit (e.g. Jones \& Forman 1984), 
while in some cases, the sizes of the excluded
regions are not at all obvious.  
The introduction of a double $\beta$ model thus provides 
a way to quantitatively describe the central excess emission. 
In fact, as will be shown in this paper, our understanding of some
properties of clusters such as the determinations of gas density and 
dynamical mass is closely connected to the issue as to whether or not 
the central narrow  component is properly taken into account. 
On the other hand, there have been increasing studies both 
theoretically and observationally on the possibility
that intracluster gas is multiphase (Nulsen 1986; White \& Fabian 1995; 
Gunn \& Thomas 1996; Nagai, Sulkanen \& Evrard 2000; Buote 2000, etc.),
among which the detection of excess of low-energy photons in X-ray
spectra  relative to the average X-ray temperature in some clusters 
gives a convincing support to the presence of a cold gas 
component either concentrated in the central core or distributed over
the entire cluster.  A double $\beta$ model may correspond to 
the simplest case for multiphase medium, and therefore a detailed 
study of its properties will be helpful for our further 
investigation of the complex and multiphase intracluster gas.

In this paper, we present the gas distribution inferred from a double
$\beta$ model (section 2.1) and the total cluster mass under the assumption 
that the two-phase gas is in hydrostatic equilibrium with the
underlying gravitational potential of the cluster (section 2.2). 
We demonstrate the difference in the cluster properties characterized by 
a single and double $\beta$ model using two strong cooling flow clusters, 
A2597 and A2390 (section 3). 
Because the sharp peak in the central X-ray emission can
be quantitatively described by the second $\beta$ model, we will be
able to estimate more accurately the cluster mass enclosed within the 
central core and compare with that derived from strong gravitational lensing.
We will examine whether the double $\beta$ model corrected X-ray mass can 
be reconciled with the strong lensing result (e.g. Wu \& Fang 1997).
Finally, as a consequence of adopting a double $\beta$ model fit,  
the $\beta$ parameters in the two components 
may both become larger than the value in a single $\beta$ model fit.
We will study the possibility of  resolving the well-known $\beta$ 
discrepancy, using a sample of 33 clusters drawn from the recent work of MME.  
Throughout this paper we assume 
$H_0=50$ km s$^{-1}$ Mpc$^{-1}$ and $\Omega_0=1$.

\section{THE MODEL}

\subsection{GAS DISTRIBUTION}

If we assume that the intracluster gas has two-phases with different 
electron temperatures $T_i$, where and also hereafter 
the subscript number ($i=1,2$) refers to the two-phases,
the X-ray emission per unit volume in an energy band from
$E_1$ to $E_2$, according to thermal bremsstrahlung, is 
%1
\begin{equation}
\frac{dL_x}{dV}= \displaystyle\sum_{i}\alpha(T_i)
                n_{ei}\displaystyle\sum_{z,Z}N_{Z,z}z^2 g_i 
\end{equation}
where
%2
\begin{equation}
\alpha(T_i)=\frac{2^4e^6}{3m_e\hbar c^2}
		\left(\frac{2\pi kT_i}{3m_ec^2}\right)^{1/2},
\end{equation}
and
%3 
\begin{eqnarray}
g_i=\int_{E_1}^{E_2}\overline{g}_{ff}(T_i,\nu)e^{-\frac{h\nu}{kT_i}}
     d\left(\frac{h\nu}{kT_i}\right), 
\end{eqnarray}
in which $n_{e}$ is the electron number density, 
$N_{Z,z}$ is the ion number density
with atomic number $Z$ and effective charges $z$, and 
$\overline{g}_{ff}$ is the Guant factor of the free-free emission.
We define the average electron weight $\mu_{ei}$ (or $\mu_e$) as
%4
\begin{equation}
\mu_{ei}n_{ei}=\mu_en_e=\displaystyle\sum_{z,Z}N_{Z,z}z^2
\end{equation}
where $n_e=\displaystyle\sum_{i} n_{ei}$ 
is the total electron number density, so that 
%5
\begin{equation}
\displaystyle\sum_{i}\frac{1}{\mu_{ei}}=\frac{1}{\mu_e}.
\end{equation}
In particular, if intracluster gas is assumed to have the cosmic primordial
abundances of hydrogen and helium, then $\mu_e=2/(1+X)$ with $X$ being 
the primordial hydrogen mass fraction $X=0.768$.

Suppose that the two-phase emission gives rise to, correspondingly, 
the two components in the observed X-ray surface brightness profile
described by a double $\beta$ model:
%6
\begin{equation}
S(r)=\displaystyle\sum_{i}S_{0i}\left(1+\frac{r^2}{r_{ci}^2}\right)
     ^{-3\beta_i+0.5}.
\end{equation}
By inverting $S(r)$ and comparing with eq.(1), we have 
(e.g. Cowie, Henriksen \& Mushotzky 1987)
%7
\begin{equation}
\mu_{ei}(0)n_{ei}^2(0)=\frac{4\pi^{1/2}}{\alpha(T_i)g_i}
	             \frac{\Gamma(3\beta_i)}{\Gamma(3\beta_i-\frac{1}{2})}
                     \frac{S_{0i}}{r_{ci}},
\end{equation}
and
%8 
\begin{equation}
\mu_{ei}n_{ei}^2=\mu_{ei}(0)n_{ei}^2(0)
           \left(1+\frac{r^2}{r_{ci}^2}\right)^{-3\beta_i}.
\end{equation}
Eliminating $\mu_{ei}$ in the above equations in terms of 
eqs.(4) and (5), we can get the electron number densities for the two  
components as well as  the combined electron number density $n_e$
%9
\begin{equation}
\frac{n_{ei}}{n_{ei}(0)}=
      \left(1+\frac{r^2}{r_{ci}^2}\right)^{\frac{-3\beta_i}{2}}
      \left[\frac{n_{e0}\left(1+\frac{r^2}{r_{ci}^2}\right)^{-3\beta_i}}
           {\displaystyle\sum_i 
           n_{ei}(0)\left(1+\frac{r^2}{r_{ci}^2}\right)^{-3\beta_i}}
      \right]^{1/2},
\end{equation}
or 
%10
\begin{equation}
n_{ei} =  \left(\frac{n_{e0}}{n_e}\right)\tilde{n}_{ei}, 
\end{equation}
and
%11
\begin{equation}
n_e=\displaystyle\sum_{i}n_{ei}=
    \left[n_{e0}\displaystyle\sum_{i}\tilde{n}_{ei}\right]^{1/2},
\end{equation}
in which
%12
\begin{equation}
\tilde{n}_{ei}\equiv n_{ei}(0)\left(1+\frac{r^2}{r_{ci}^2}\right)^{-3\beta_i}.
\end{equation}
Note that although eq.(11) can be formally written as 
$n_{gas}=[n_{gas1}^2+n_{gas2}^2]^{1/2}$, which has been adopted in some 
recent work (e.g. Reiprich \& B\"ohringer 1999), the determination of 
the central gas density for each component ($n_{gas1}$ or $n_{gas2}$)
described by a single $\beta$ model is dependent on both components 
of $S(r)$. In terms of eq.(7), 
the central electron number density, $n_{ei}(0)$, 
is related to the observationally fitted
central surface brightness $S_{0i}$  through
%13
\begin{equation}
n^2_{ei}(0)=\frac{4\pi^{1/2}}{\alpha(T_i)g_i\mu_e}
	             \frac{\Gamma(3\beta_i)}{\Gamma(3\beta_i-\frac{1}{2})}
                     \frac{S_{0i}}{r_{ci}} A_{ij},
\end{equation}
in which
%14
\begin{equation}
\frac{1}{A_{ij}}= 1+\frac{S_{0j}}{S_{0i}}\frac{r_{ci}}{r_{cj}}
                     \frac{g_i}{g_j}
                     \left(\frac{T_i}{T_j}\right)^{1/2}
                     \frac{\Gamma(3\beta_j)}{\Gamma(3\beta_i)}
		     \frac{\Gamma(3\beta_i-\frac{1}{2})}
                          {\Gamma(3\beta_j-\frac{1}{2})},
\end{equation}
where and also hereafter $j=1,2$ and $j\neq i$.
Of course, one can also estimate $n_{ei}(0)$ using the total X-ray 
luminosity $L_x$ of the cluster (eq.[1]). For example, in the case of 
isothermal gas distributions for the two-phase gas we have
%15
\begin{equation}
n^2_{ei}(0)=\frac{L_xA_{ij}B_{ij}}
		 {\pi^{3/2}\alpha(T_i)g_i\mu_er_{ci}^3}
	             \frac{\Gamma(3\beta_i)}{\Gamma(3\beta_i-\frac{3}{2})},
\end{equation}
where
%16
\begin{equation}
\frac{1}{B_{ij}}=1+\frac{S_{0j}}{S_{0i}}\frac{r_{cj}^2}{r_{ci}^2}
                     \frac{3\beta_i-\frac{3}{2}}{3\beta_j-\frac{3}{2}}.\\
\end{equation}
Apparently, the total X-ray luminosity can be estimated simply from
%17
\begin{equation}
L_x=\displaystyle\sum_{i}\frac{4\pi^2 r_{ci}^2 S_{0i}}{3\beta_i-\frac{3}{2}}.
\end{equation}

\subsection{CLUSTER MASS}

The total mass in gas within a sphere of radius $r$ is 
%18
\begin{equation}
M_{gas}(r)=4\pi\mu_e m_p n_{e0}
         \int_0^r
         \left[\frac{1}{n_{e0}}\displaystyle\sum_{i}
                \tilde{n}_{ei}(r)\right]^{1/2}
          r^2dr,
\end{equation}
If the X-ray emitting
gas is in hydrostatic equilibrium with the underlying gravitational
potential of the cluster, the total dynamical mass within $r$ is
%19
\begin{equation}
M(r)=-\frac{r^2}{G\mu_em_p n_e}
     \displaystyle\sum_i \frac{d(n_ikT_i)}{dr},
\end{equation}
in which $n_i$ is the total particle number density for
the $i$th-phase gas. If the abundances of hydrogen
and helium for a single-phase gas are assumed to remain unchanged over 
the whole cluster, we have $n_i=\mu_{si}n_{ei}$,
where $\mu_{si}$ is a proportionality coefficient.  
If the two-phases both have the cosmic mixed-abundances of hydrogen
and helium, $\mu_{si}=1.934$. Next, we 
assume a polytropic equation of state for each single-phase gas,
namely, $T_i=T_{i0}[n_{ei}/n_{ei}(0)]^{\gamma_i-1}$. In this case,  
eq.(19) can be written to be
%20
\begin{equation}
M(r)=  -\displaystyle\sum_i \frac{\gamma_i kT_{i0}r}{G\mu_im_p}
        \left(\frac{n_{ei}}{n_{ei}(0)}\right)
	\left(\frac{n_{ei}}{n_e}\right)
        \left(\frac{d\ln n_{ei}}{d\ln r}\right),
\end{equation}
where $\mu_i=\mu_e/\mu_{si}$. 
A straightforward computation using the electron number density found
in the above subsection yields
%21
\begin{eqnarray}
M(r)= & \displaystyle\sum_{i,j} M_i(r)\left(\frac{n_{ei}}{n_e}\right)
        \;\cdot\;    \nonumber\\
      &  \left[1+\left(1-\frac{\beta_j}{\beta_i}
                \frac{r^2+r_{ci}^2}{r^2+r_{cj}^2}\right)
               \frac{\tilde{n}_{ej}}{\tilde{n}_{ei}+\tilde{n}_{ej}}
        \right], 
\end{eqnarray}
where $M_i(r)$ is the total cluster mass determined
by assuming that the X-ray emission is from a single-phase medium
described by a single $\beta$ model:
%22
\begin{equation}
M_i(r)=3\beta_i\gamma_i\frac{kT_{i0}r}{G\mu_im_p}
         \frac{r^2}{r^2+r_{ci}^2}
          \left(\frac{n_{ei}(r)}{n_{ei}(0)}\right)^{\gamma_i-1}.
\end{equation}
It is easy to show that eq.(21) reduces to the form of eq.(22)
if the two-phase gas components are assumed to be identical
or the second component vanishes.

\section{APPLICATION TO X-RAY CLUSTERS}

\subsection{A2579}

We first choose a typical cooling flow cluster, A2579, to demonstrate 
how a double $\beta$ model works.  A2579 has been extensively
studied by Sarazin \& McNamara (1997). They showed that 
the merged ROSAT HRI and PSPC surface
brightness profile $S(x)$ cannot be fitted by a single $\beta$ model unless 
the data points within the central region of radius of 0.18 Mpc are removed.
Meanwhile,  the core radius has not been well constrained 
due to the presence of the centrally peaked X-ray emission, and 
the gas and total masses within the central region derived from 
the deconvolved gas density are apparently 
larger than those obtained from the best-fit single $\beta$ model. 
We now  make an attempt to fit the same data set $S(x)$ using 
a double $\beta$ model (eq.[6]). Fig.1 displays the observed and our
best-fit X-ray surface brightness profiles of A2579.
It turns out that the goodness of the fit has been significantly improved 
when a double $\beta$ model is applied to the entire data points,
yielding  $\chi^2/\nu=41.16/26$. Recall that 
the minimum  $\chi^2/\nu$ is $59.99$ for 29 d.o.f in a single $\beta$ model 
fitting (Sarazin \& McNamara 1997). 
The best-fit values of the $\beta$ parameter and core radius for the 
narrow and extended components are 
 $(\beta_1,r_{c1})=(0.70\pm0.32,0.047\pm0.016$ Mpc) and 
 $(\beta_2,r_{c2})=(0.66\pm0.03,0.15\pm0.05$ Mpc), respectively. 
Note that our fitted $\beta$ parameter for the extended component 
is close to the value ($0.64$) obtained by  Sarazin \& McNamara (1997) 
who have excluded the central cooling flow region. 
Alternatively, the two components in our double $\beta$ model
exhibit similar $\beta$ values. The same
situation has been known for A1795 (Xu et al. 1998), while MME 
assumed the two components of a double  $\beta$ model to have the same
$\beta$ in their analysis of 45 nearby clusters. 
We have also tried the double $\beta$ fit by correcting the PSF of
the ROSAT HRI with a FWHM of 4 arcseconds. This results in
a slightly smaller core radius for the narrow component
$(\beta_1,r_{c1})=(0.67\pm0.33,0.041\pm0.016$ Mpc), while
the extended component remains almost unchanged. 
Finally, we have compared in Fig.1 the electron number density given 
by the deprojection technique with that calculated 
from our best-fit double $\beta$ model  assuming 
$T_1=2.0$ keV and $T_2=4.34$ keV for the narrow and extended components,
respectively. The agreement between the two results is good  
with $\chi^2=41.06$ for 32 d.o.f. although 
the very central point shows a deviation from the expectation of
our double $\beta$ model.  Note that the temperature of the cluster,
especially the outer cluster, is poorly constrained. Our choice of
$T_1$ and $T_2$ is simply based on the single temperature 
model fits by Sarazin \& McNamara (1997) 
to the ROSAT PSPC spectra of the core region ($0.1$ Mpc) 
and the outer cluster ($0.25$--$1$ Mpc), respectively.

\subsection{A2390}

The reason why we choose A2390 for a further demonstration of the double
$\beta$ model is as follows: First, a single $\beta$ model fails to fit 
the merged ROSAT HRI and PSPC surface brightness profile $S(x)$ of 
the cluster, which has been well observed out to a radius of $\sim2$ Mpc 
(B\"ohringer et al. 1998); Second, the mass distribution of the cluster
has been mapped with the weak lensing technique (Squires et al. 1996).
It deserves to be examined whether the cluster mass derived from  
a double $\beta$ model can be reconciled with the weak lensing result.
The latter is believed to provide a reliable mass estimate, independently
of the cluster matter contents and their dynamical states.
Third, none of the suggested models in literature 
based on the fitting of $S_x$ have 
reproduced the mass profile given by the weak lensing analysis
(B\"ohringer et al. 1998).  In Fig.2 we plot the X-ray
surface brightness profile of A2390 observed from the ROSAT HRI and PSPC 
(B\"ohringer et al. 1998), along with a single and double $\beta$ model
fitting. The best-fit parameters are
$(\beta,r_c)=(0.56\pm0.01,0.11\pm0.01$ Mpc) for a single $\beta$ model
with  $\chi^2/\nu=209.08/40$, and 
$(\beta_1,r_{c1})=(0.92\pm0.22,0.46\pm0.16$ Mpc) and 
$(\beta_2,r_{c2})=(0.51\pm0.01,0.046\pm0.010$ Mpc) for a double
$\beta$ model with  $\chi^2/\nu=52.48/37$, respectively. 
Indeed, a single $\beta$ model fit is not acceptable.
B\"ohringer et al. (1998) obtained a good fit by 
introducing a Gaussian peak plus a single $\beta$ model. 
Our double $\beta$ model does provide a significantly reduced $\chi^2$ 
fit to the entire data points although the two components 
appear to be unusual in shape.  
Because A2390 is located at moderate redshift ($z=0.228$), 
we may need to consider whether the core radius 
is overestimated in the above fitting due to the PSF of the ROSAT HRI.
We repeat the double $\beta$ model fitting using 
a PSF with a FWHM of 4 arcseconds. The best fit parameters become 
$(\beta_1,r_{c1})=(0.98\pm0.12,0.49\pm0.08$ Mpc) and 
$(\beta_2,r_{c2})=(0.50\pm0.01,0.035\pm0.005$ Mpc).  
Indeed, for the second component the core radius is apparently reduced,
while the $\beta$ parameter remains roughly the same.

Now, we calculate the projected cluster
mass within  radius $r$. The major uncertainty here is the X-ray temperatures
for the two components of the double $\beta$ model. The presence of the
cooling flow may lead to an underestimate of 
the overall temperature of the cluster.
The reported temperature of the cluster in the literature ranges from 
the average value $7.7_{-0.98}^{+1.29}$ keV over the cluster 
(Rizza et al. 1998)  to the cooling flow corrected value
$23.09\pm35.12$ keV (White 2000). Moreover, there is an apparent discrepancy
between the ROSAT PSPC and ASCA GIS measured temperatures within 
the central region of the cluster.
In the following  we first adopt the two-temperature model of
B\"ohringer et al. (1998), which consists of a narrow,
low-temperature component with $T_1$ and an extended, high-temperature 
one with $T_2$. The projected cluster mass deriving from this double
$\beta$ model based on the best-fit values of $T_1=2.0$ keV and
$T_2=11.5$ keV by B\"ohringer et al. (1998)
is shown in Fig.2. While the shape of the resulting mass profile
seems to be consistent with the weak lensing result,
the double $\beta$ model with the above temperature data has significantly
underestimated the cluster mass as compared with the weak lensing 
analysis. Next, we treat the two temperatures $T_1$ and $T_2$ 
as free parameters and search for ($T_1$, $T_2$) 
that give rise to the best fit of the derived cluster mass from
the double $\beta$ model to the weak lensing result.
This yields $T_1=4.32$ keV and $T_2=20.0$ keV for the narrow and extended  
components, respectively, with a reduced $\chi^2/\nu$ value
of $2.38$. Interestingly, these best-fit temperatures for the narrow
and extended components are consistent with the reported value 
of $\sim4$ keV in the central region of the cluster inside $\sim 0.5$ Mpc 
(B\"ohringer et al. 1998) and the cooling flow corrected value
$23.09\pm35.12$ keV (White 2000), respectively.  
We have also displayed in Fig.2 the projected mass profiles derived from 
other two models for $S_x$: (1)a single $\beta$ model and 
(2)a single $\beta$ model plus a Gaussian peak.
For these two models we use an isothermal plasma of $T=9.0$ keV,
a value obtained from the ASCA spectra within a radius
of $1.7$ Mpc (B\"ohringer et al. 1998). 
The goodness of the fits of the resulting cluster masses 
to the weak lensing measurement reads 
$\chi^2/\nu=3.55$ and $4.63$ for  models (1) and (2), respectively. 
In a similar way to the above analysis for the double $\beta$ model,
we have performed the $\chi^2$ fits of the projected cluster masses 
given by models (1) and (2) to the weak lensing determined mass profile
by treating the temperature as a free parameter. 
It turns out that the best-fit values of temperature for the two models
are $T=8.74$ keV with $\chi^2/\nu=3.48$ 
and $T=9.15$ keV with $\chi^2/\nu=4.59$, respectively.
These values are well within the ROSAT and ASCA spectral measurements.
It is likely that 
the double $\beta$ model fit gives the smallest $\chi^2$ value, and thus
provides the most precise description for the intracluster gas,
although the current fit is still unsatisfactory. 
Nevertheless, it should be pointed out
that the measurement uncertainties in the fitting of $S_x$ and 
the spectral measurement of temperature have not been included
in the final mass estimate of the cluster. In particular,
using a double $\beta$ model instead of a single $\beta$ model 
will lead to a large uncertainty in the derived cluster mass
since the central surface brightness, core radius and $\beta$
parameter in a double $\beta$ model fit
cannot be well constrained at present.
Finally,  as compared with the single $\beta$ model, 
the double $\beta$ model results in a significantly large
cluster mass within small radii ($\sim100$ kpc), and the overall 
dark matter distribution clearly exhibits two distinct length scales. 
Similar result has also been reported in previous studies 
(e.g. Ikebe et al. 1996). However, it is unlikely that 
the central mass excess as a result of the employment
of the double $\beta$ model can resolve the mass discrepancy between
the strong lensing and X-ray measurements  
(Wu \& Fang 1997), despite the fact that our fitted core radius 
($r_{c2}=0.035\pm0.005$ Mpc) is relatively small.
Our result agrees with the recent analysis by Lewis et al. (1999), 
who found the X-ray mass within the arc radius is $2.3$ times smaller 
than the value reported by Allen (1998),  $2.1\times10^{14}M_{\odot}$.
This indicates again that the previously
claimed mass discrepancy between strong lensing and other methods 
has probably arisen from the oversimplification of lensing model for 
matter distribution in the central cores of the strong lensing clusters.

\section{The $\beta$ DISCREPANCY}

In this section we discuss whether the well-known puzzle,
the so-called $\beta$ discrepancy (e.g. Bahcall \& Lubin 1994), 
may have also arisen from the 
employment of the conventional $\beta$ model for intracluster gas.
Briefly, if both galaxies and gas are the tracers of the depth and shape
of a common gravitational potential of a cluster, we would expect
the following identity (Cavaliere \& Fusco-Femiano 1976; 
Bahcall \& Lubin 1994)
%23
\begin{equation}
\beta_{spec}\equiv \frac{\sigma^2}{kT/\mu m_p}=
    \frac{d\ln n_{gas}/d\ln r}{d\ln n_{gal}/d\ln r}\equiv \beta_{fit},
\end{equation}
where $\sigma$ is the velocity dispersion of galaxies,
and the number densities of galaxies and gas are 
denoted by $n_{gal}$ and $n_{gas}$, respectively. 
For simplicity, we have assumed an isotropic and isothermal matter
distribution for the cluster. The $\beta$
discrepancy arises from a statistical estimate of  $\beta_{spec}$ 
and $\beta_{fit}$, which turns out that 
on average $\beta_{spec}$ ($\approx1$)  is larger than 
$\beta_{fit}$ ($\approx2/3$). 
Basically, two solutions to the puzzle have so far been suggested: 
Either the galaxy profile in clusters may be less steeper than 
the simplified King model, or 
the present X-ray and optical observations have only probed  some 
finite regions of clusters so that the asymptotic value of $\beta_{fit}$ 
has not yet been reached (Gerbal, Durret \& Lachi\`eze-Rey 1994).

Numerical studies of cluster formation and
evolution (e.g. Navarro, Frenk \& White 1995;   Lewis et al. 1999)
have shown that the gas density profiles of clusters become steeper 
with radius and therefore, a larger $\beta$ value
can be obtained if the fit to the observed X-ray surface
brightness profile is extended to large radii.
Indeed,  Vikhlinin et al.(1999) have tried a single $\beta$ model fit 
to the gas distribution in the outer regions of clusters 
between 0.3 and 1 virial radius. 
They found that the typical $\beta$ values among the 39 clusters 
observed by ROSAT PSPC would be larger by $\approx0.05$ than that 
derived from the global fit. These 
results provide a useful clue to our reconsideration
of the $\beta$ discrepancy because only the asymptotic $\beta$ value
at large radii is of significance for the evaluation
of $\beta_{fit}$.  As has been pointed out in the above section,
the $\beta$ parameters for the two components of a double $\beta$ model 
may both become larger than the value in a single $\beta$ model fit. 
Replacing the $\beta$ in eq.(23) by the $\beta$ parameter of the extended 
component of a double $\beta$ model 
may give rise to a larger value of  $\beta_{fit}$. Hence, 
this opens a new possibility of reconciling  $\beta_{spec}$ with 
$\beta_{fit}$.

\subsection{CLUSTER SAMPLE}

We begin with the sample of 45 ROSAT PSPC observed nearby clusters 
compiled recently by MME.
We choose this sample because MME have already 
tried a double $\beta$ model fit to 
the 18 clusters of their sample, 
which can be used for the purpose of comparison,  
although their double $\beta$ model assumed the same $\beta$ parameter 
for the two components. In order to estimate $\beta_{spec}$, 
we exclude those clusters whose velocity dispersion and temperature are not 
observationally determined. This reduces the MME sample to 33 clusters
(see Table 1 and Table 2).  
We reanalyze the archival ROSAT PSPC 
imaging data of the 33 clusters following the same procedure of MME,
who binned the X-ray surface brightness according to photon counts.
We have also tried to measure $S_x(r)$ using the
concentric rings of equal width ($1$ arcminute) 
but found that the data points are too few for some clusters.
To facilitate the double $\beta$ model fit to $S_x(r)$, we will use the
results of $S_x(r)$ from MME.

We fit the observed X-ray surface brightness of clusters to a double
$\beta$ model described by eq.(6).  
We employ the Monte Carlo simulations and the $\chi^2$-fit to
obtain the best-fit parameters ($S_{0i}$, $r_{ci}$, $\beta_i$) and 
estimate their error bars, in which we keep the same outer radii 
of the fitting regions as those defined by MME.
Essentially, the fitting can be classified as two types, and 
a typical example for each type is displayed in Fig.3.
In the first case which is applied to 15 clusters in Table 1, 
the narrow component contributes mainly to the central X-ray emission and 
exhibits a sharp drop at large radii, while the extended component
dominates the emission in the outer region. 
As for our example A2052 in Fig.3,  a single $\beta$ model fit gives
$\beta=0.593\pm0.050$ but the minimum $\chi^2$ is 310.67 for
49 d.o.f. Our double $\beta$ fit yields $\beta_1=0.810\pm0.194$ and
$\beta_2=0.773\pm0.029$ with $\chi^2/\nu=60.67/46$. 
It appears that our best-fit $\beta$ parameter
for the extended component is comparable to the one ($0.712$) 
found by MME using a double $\beta$ model fit with the same 
$\beta$ parameter for the two components.  
In the second case, the central emission is similarly governed 
by the narrow component. However, the narrow component exhibits
a flatter slope than the extended component does in the outer region.
Yet, the amplitude of the narrow component is significantly
smaller than that of the extended component. We will thus 
take a less vigorous approach to the selection of the 
$\beta$ parameter: The value of the extended component will be
used in our evaluation of $\beta_{fit}$ below.  
Mathematically, the fitted X-ray surface brightness profile will be
eventually dominated by the narrow and shallower component at very large radii
because of its smaller value of $\beta$.
For A1060 shown in Fig.3, the minimum $\chi^2$ of $598.3$ in a 
single $\beta$ model fit for 113 d.o.f has been reduced to $139.9$ in our
double $\beta$ model fit for 110 d.o.f.  The best-fit $\beta$
parameter of the extended component is  $0.732\pm0.065$, 
in comparison with the MME result $\beta=0.703_{-0.036}^{+0.044}$.

The best-fit $\beta$ parameters by a double $\beta$ model  
for our cluster sample are listed in Table 1, 
together with the results for a single $\beta$ model fit,
Quoted errors are $68\%$ confidence limits.
Note that the $\beta$ values of the narrow components for
about 1/3 of our clusters have not been well constrained. This is partially
due to the fact that a single $\beta$ model fit is acceptable for
some clusters such as A401, A1367, A1651 and A4059. Another reason comes
from the sparse data points at small radii, which makes it difficult to 
place a robust constraint on the narrow component.

\begin{table*}
\vskip 0.2truein
\begin{center}
\caption{Cluster Sample:  $\beta_{fit}$}
%\begin{scriptsize}
\vskip 0.2truein
\begin{tabular}{ l l l l l l }
\hline
cluster &  $\beta^a$ & $\beta_1^b$ & $\beta_2^c$ & $\beta_{MME}^d$ \\
\hline
A85	& $0.579\pm0.009$  & $0.975\pm0.305$  & $0.687\pm0.010$ & 0.662 \\
A262  	& $0.447\pm0.004$  & $0.542\pm0.169$  & $0.575\pm0.045$ & 0.556 \\
A401	& $0.612\pm0.006$  & $1.016\pm1.05$   & $0.679\pm0.040$ & 0.606 \\
A426	& $0.548\pm0.007$  & $0.688\pm0.045$  & $1.156\pm0.132$ & 0.748 \\
A478	& $0.661\pm0.008$  & $0.723\pm0.242$  & $0.736\pm0.041$ & 0.713 \\
A496	& $0.544\pm0.008$  & $0.734\pm0.218$  & $0.708\pm0.034$ & 0.650 \\
A754	& $0.761\pm0.025$  & $1.01\pm0.26$    & $1.60\pm0.11$   & 0.614 \\
A1060	& $0.587\pm0.012$  & $0.601\pm0.326$  & $0.732\pm0.065$ & 0.703 \\
A1367 	& $0.766\pm0.023$  & $0.785\pm0.671$  & $1.554\pm0.731$ & 0.707 \\
A1651	& $0.637\pm0.013$  & $1.22\pm1.76$    & $0.705\pm0.019$ & 0.616 \\
A1656	& $0.683\pm0.048$  & $0.504\pm0.090$  & $0.851\pm0.131$ & 0.705 \\ 
A1689	& $0.752\pm0.026$  & $1.25\pm0.07$    & $0.954\pm0.040$ & 0.648 \\ 
A1795 	& $0.652\pm0.010$  & $1.01\pm0.146$   & $0.691\pm0.040$ & 0.790 \\
A2029	& $0.637\pm0.010$  & $0.746\pm0.178$  & $0.833\pm0.071$ & 0.705 \\
A2052	& $0.564\pm0.005$  & $0.810\pm0.194$  & $0.773\pm0.029$ & 0.712 \\
A2063	& $0.576\pm0.010$  & $1.10\pm0.97$    & $0.658\pm0.014$ & 0.706 \\ 
A2142	& $0.656\pm0.010$  & $1.305\pm1.139$  & $0.786\pm0.037$ & 0.787 \\
A2199 	& $0.611\pm0.003$  & $0.631\pm0.209$  & $0.679\pm0.040$ & 0.663 \\
A2244 	& $0.580\pm0.017$  & $0.846\pm0.617$  & $0.644\pm0.283$ & 0.594 \\
A2256	& $0.822\pm0.016$  & $0.348\pm0.030$  & $0.926\pm0.031$ & 0.792 \\
A2319   & $0.578\pm0.009$  & $0.840\pm0.995$  & $0.745\pm0.059$ & 0.828 \\
A3112 	& $0.568\pm0.003$  & $0.735\pm0.155$  & $0.535\pm0.014$ & 0.562 \\
A3266	& $0.795\pm0.024$  & $2.09\pm3.17$    & $1.07\pm0.05$   & 0.744 \\
A3391	& $0.510\pm0.016$  & $0.355\pm0.038$  & $0.704\pm0.275$ & 0.541 \\
A3526	& $0.464\pm0.006$  & $0.531\pm0.051$  & $0.824\pm0.169$ & 0.569 \\
A3532	& $0.598\pm0.023$  & $0.471\pm0.233$  & $0.783\pm0.425$ & 0.589 \\
A3562 	& $0.488\pm0.004$  & $0.610\pm0.578$  & $0.551\pm0.039$ & 0.470 \\ 
A3571	& $0.645\pm0.011$  & $0.822\pm0.652$  & $0.795\pm0.061$ & 0.610 \\
A3667 	& $0.598\pm0.014$  & $0.864\pm0.317$  & $1.26\pm0.35$   & 0.541 \\
A4059	& $0.603\pm0.013$  & $0.790\pm0.734$  & $0.740\pm0.033$ & 0.558 \\
AWM7	& $0.600\pm0.013$  & $0.518\pm0.058$  & $0.801\pm0.070$ & 0.678 \\
Cygnus-A& $0.470\pm0.006$  & $0.587\pm0.048$  & $0.902\pm0.618$ & 0.472 \\
MKW3s 	& $0.598\pm0.009$  & $1.16\pm1.79$    & $0.645\pm0.017$ & 0.562 \\
%\hline
	&	           &                  &                 & \\
Average & $0.613\pm0.090$  & $0.819\pm0.400$  & $0.833\pm0.334$ & 0.648 \\
 \hline                                                             
\end{tabular}
%\end{scriptsize}
\end{center}
\parbox {3.5in}{$^a$The single $\beta$ model;}\\
\parbox {3.5in}{$^b$Narrow component in the double $\beta$ model;}\\
\parbox {3.5in}{$^c$Extended component in the double $\beta$ model;}
\parbox {3.5in}{$^d$The best-fit $\beta$ parameter by MME. Error bars
		are not listed.}
 \end{table*}

We searched the literature for the velocity dispersion of galaxies and
the temperature of X-ray emitting gas in our cluster sample (Table 2).
Essentially, we take the X-ray temperature data of  White (2000)
(except A2244 and A3532)
because the effect of the cooling flow on the cluster temperature 
has been corrected for. The majority of the velocity dispersion data 
are chosen from the substructure corrected values by Girardi et al. (1998).
The $\beta_{spec}$ value (the significance at $68\%$ confidence) 
is correspondingly calculated 
for each cluster and listed in Table 2.

\begin{table*}
\vskip 0.2truein
\begin{center}
\caption{Cluster Sample: $\beta_{spec}$}
%\begin{scriptsize}
\vskip 0.2truein
\begin{tabular}{ l l l l l  }
\hline
cluster &  $\sigma$ (km s$^{-1}$) & $T$ (keV) & $\beta_{spec}$	\\
\hline
A85	& $969^{+95}_{-61}$  a  & $6.74^{+0.50}_{-0.50}$ b	
	& $0.850^{+0.257}_{-0.155}$    \\
A262	& $525^{+47}_{-33}$  a  & $2.29^{+0.12}_{-0.09}$ b
	& $0.734^{+0.173}_{-0.121}$    \\
A401	& $1152^{+86}_{-70}$ a  & $10.68^{+1.11}_{-0.94}$ b    
	& $0.758^{+0.202}_{-0.152}$    \\
A426	& $1026^{+106}_{-64}$ a & $7.71^{+0.29}_{-0.37}$ b   
 	& $0.833^{+0.232}_{-0.127}$    \\
A478	& $904^{+261}_{-140}$ c & $7.42^{+0.71}_{-0.54}$ b  
	& $0.672^{+0.531}_{-0.234}$    \\
A496	& $687^{+89}_{-76}$ a   & $4.51^{+0.17}_{-0.15}$ b 
	& $0.638^{+0.204}_{-0.152}$    \\
A754	& $662^{+77}_{-50}$ a & $12.85^{+1.77}_{-1.35}$ b    
	& $0.208^{+0.082}_{-0.052}$    \\
A1060	& $610^{+52}_{-43}$ a   & $3.27^{+0.11}_{-0.09}$ b 
	& $0.694^{+0.146}_{-0.114}$   \\
A1367	& $822^{+88}_{-72}$ c   & $3.99^{+0.48}_{-0.48}$ b
	& $1.033^{+0.406}_{-0.265}$ \\
A1651	& $965^{+160}_{-107}$ c & $7.15^{+0.84}_{-0.62}$ b   
	& $0.794^{+0.388}_{-0.232}$   \\
A1656	& $821^{+49}_{-38}$ c  & $10.03^{+0.89}_{-0.81}$ b  
	& $0.410^{+0.091}_{-0.067}$    \\ 
A1689	& $1470^{+210}_{-160}$ d  & $9.48^{+1.36}_{-0.52}$ b  
	& $1.3901^{+0.531}_{-0.425}$    \\
A1795	& $834^{+85}_{-76}$ a   & $7.26^{+0.51}_{-0.40}$ b 
	& $0.584^{+0.167}_{-0.133}$    \\
A2029	& $1164^{+98}_{-78}$ a  & $8.22^{+0.58}_{-0.20}$ b 
	& $1.005^{+0.206}_{-0.188}$   \\
A2052	& $714^{+143}_{-148}$ e   & $3.30^{+0.16}_{-0.13}$ b   
	& $0.942^{+0.471}_{-0.377}$   \\
A2063	& $667^{+55}_{-41}$ a  & $3.90^{+0.51}_{-0.38}$ b 
	& $0.696^{+0.207}_{-0.154}$   \\ 
A2142	& $1132^{+110}_{-92}$ a & $10.96^{+2.56}_{-1.58}$ b   
	& $0.713^{+0.290}_{-0.225}$   \\
A2199	& $801^{+92}_{-61}$  a  & $4.70^{+0.13}_{-0.15}$ b  
	& $0.833^{+0.236}_{-0.141}$   \\
A2244	& $1204^{+232}_{-232}$ f & $8.47^{+0.43}_{-0.42}$ g   
	& $1.044^{+0.518}_{-0.396}$      \\
A2256	& $1348^{+86}_{-64}$ a  & $8.69^{+1.06}_{-1.06}$ b   
	& $1.275^{+0.368}_{-0.244}$      \\
A2319	& $1545^{+95}_{-77}$ a  & $13.60^{+2.22}_{-2.22}$ b
	& $1.070^{+0.371}_{-0.240}$  \\  
A3112	& $552^{+86}_{-63}$ h   & $4.69^{+0.27}_{-0.26}$ b 
	& $0.396^{+0.164}_{-0.102}$   \\
A3266	& $1107^{+82}_{-65}$ a  & $9.69^{+0.97}_{-0.92}$ b 
	& $0.771^{+0.212}_{-0.150}$    \\
A3391	& $663^{+195}_{-112}$ a   & $6.90^{+1.47}_{-0.86}$ b  
	& $0.389^{+0.355}_{-0.167}$    \\
A3526	& $780^{+100}_{-100}$ e   & $4.04^{+0.11}_{-0.11}$ b 
	& $0.918^{+0.283}_{-0.239}$    \\
A3532	& $738^{+112}_{-85}$ a  & $4.4^{+4.7}_{-1.3}$ i  
	& $0.755^{+0.667}_{-0.469}$    \\
A3562	& $736^{+49}_{-36}$ a   & $6.96^{+1.77}_{-0.95}$ b 
	& $0.475^{+0.151}_{-0.132}$    \\ 
A3571	& $1045^{+109}_{-90}$ a & $8.12^{+0.42}_{-0.39}$ b   
	& $0.820^{+0.230}_{-0.169}$    \\
A3667	& $1045^{+62}_{-47}$ a  & $8.11^{+0.82}_{-0.73}$ b
	& $0.709^{+0.172}_{-0.126}$    \\
A4059	& $845^{+280}_{-140}$ j & $4.05^{+0.23}_{-0.19}$ b
	& $1.075^{+0.924}_{-0.367}$       \\
AWM7	& $864^{+110}_{-80}$ a & $3.96^{+0.16}_{-0.14}$ b 
	& $1.150^{+0.365}_{-0.240}$       \\
Cygnus-A& $1581^{+286}_{-197}$ k & $39.40^{+2.66}_{-2.66}$ b 
	& $0.387^{+0.192}_{-0.109}$       \\
MKW3s	& $610^{+69}_{-52}$ a   & $3.71^{+0.16}_{-0.19}$ b 
	& $0.612^{+0.187}_{-0.121}$    \\
%\hline
	&                      &        & \\
Average & $925\pm281$          & $7.86\pm6.36$
	& $0.777\pm0.367$    \\
%        &                      &       
%        & $0.809\pm0.376^*$  \\
%	&                      &        & \\
\hline
\end{tabular}
%\end{scriptsize}
\end{center}
%\parbox {3.5in}{$^*$The mean value without inclusion of A754.}\\
\parbox {3.5in}{References.--
	(a)Girardi et al. 1998;
	(b)White 2000;
	(c)Zabludoff et al. 1990;
	(d)Jones \& Forman 1999; 
	(e)Bird et al. 1995; 
	(f)Struble \& Rood  1991;
	(g)Mushotzky \& Scharf  1997;
	(h)Fadda et al. 1996;
	(i)Edge et al. 1990; 
	(j)Green et al. 1988; 
	(k)Owen et al. 1997.}\\
 \end{table*}

\subsection{THE $\beta$ PARAMETERS}

The mean value of $\beta_{spec}$ over 33 clusters in Table 2 is
$\langle\beta_{spec}\rangle=0.78\pm0.37$, which 
is consistent with previous findings within uncertainty  
[see Wu, Fang \& Xu (1998) for a summary]. This value  
can be slightly raised to $\langle\beta_{spec}\rangle=0.80\pm0.36$
if A754 is excluded from the list. This cluster gives rise to the 
smallest value of $\beta_{spec}$ because of its relatively small
velocity dispersion or high temperature. 
Similar effect was also noticed in previous work, i.e.,  
the  mean value of $\beta_{spec}$ can be biased high or low 
because of a few clusters in the sample. 
For instance, Bird, Mushotzky \& Metzler (1995) found that 
their mean value of  $\beta_{spec}=1.20$ drops to 1.09 if A2052 is
removed from their cluster list. Using a large sample of 149 clusters, 
Wu, Fang \& Xu (1998) showed  
$\langle\beta_{spec}\rangle=1.00\pm0.49$ but the median 
value is only $\langle\beta_{spec}\rangle=0.80^{+1.04}_{-0.52}$.

The mean values of $\beta_{fit}$ (we will assume 
a simplified King model for galaxy number density profile) from Table 1 are
$\langle\beta_{fit}\rangle=0.61\pm0.09$ and 
$\langle\beta_{fit}\rangle=0.83\pm0.33$ for a single and double
$\beta$ fit, respectively. 
While the error bar is still large, the value of $\langle\beta_{fit}\rangle$
has been apparently increased in the case of a
double $\beta$ model fit to $S_x(r)$. For comparison, 
the mean value of $\beta_{fit}$ in the MME sample is 
$\langle\beta_{fit}\rangle=0.61$ for a single $\beta$ model fit  
among 27 clusters, while $\langle\beta_{fit}\rangle=0.70$ 
for a double $\beta$ model fit for the rest 18 clusters. Recall that their
double $\beta$ model assumed the same $\beta$ parameter for the two 
components. Our result of $\beta_{fit}$ 
for a single $\beta$ model is in good agreement
with the one found by  MME. However, our double $\beta$  model fit yields
a slightly larger mean value of $\beta_{fit}$, which 
is nevertheless consistent with the claim ($0.84\pm0.1$) by
Bahcall \& Lubin (1994) (Note the different definition of $\beta_{fit}$),
and in particular with the numerical
result ($0.82\pm0.06$) by Navarro et al. (1995).

We now examine whether the observationally fitted parameter 
$\beta_{fit}$ in the outer regions of clusters have any 
dependence on temperature or velocity dispersion. 
Fig.4 shows the best-fit $\beta_{fit}$ versus $T$ for our cluster
sample. We fit the data points to an expression 
$\beta_{fit}=(T/{\rm keV})^{a}+b$, and find 
$(a,b)=(0.17\pm0.03,-0.73\pm0.05)$ 
and $(a,b)=(0.22\pm0.03,-0.71\pm0.08)$ for a single and double $\beta$ 
model fit, respectively.
We thus confirm the claim by  Vikhlinin et al. (1999) that 
$\beta_{fit}$ has a mild trend to increase with temperature,
although we get 
somewhat a larger variation of $\beta_{fit}$ 
from  $\beta_{fit}\approx0.56$ (0.48) for $T=3$ keV to $\approx0.93$ (0.75)
for $T=10$ keV in a double (single) model fit.

We then study the distribution of  $\beta_{spec}/\beta_{fit}$ for
each cluster, aiming at examining whether the spectroscopic parameter 
$\beta_{spec}$ correlates with the observationally fitted parameter 
$\beta_{fit}$ from $S_x(r)$. To this end, we display in Fig.5 
$\beta_{spec}$ against $\beta_{fit}$ for our 33 clusters. 
Indeed,  there is a positive correlation between
the two $\beta$ parameters. 
A $\chi^2$ fit excluding the data point of A754 yields 
$\beta_{spec}=10^{0.018\pm0.044}\beta_{fit}^{1.12\pm0.29}$,
indicating that a high $\beta_{fit}$ corresponds  to 
a high $\beta_{spec}$. Moreover,  the above relation is also consistent with 
$\beta_{spec}=\beta_{fit}$, although the mean values of 
$\beta_{spec}$ and $\beta_{fit}$
are on average smaller than unity (see below).

Finally, we examine whether there is a $\beta$ discrepancy for
clusters. we plot in Fig.6 the ratio of $\beta_{spec}$ to $\beta_{fit}$ 
for each  cluster in our cluster sample.  
For a single $\beta$ model, the mean
ratio is $\langle\beta_{spec}/\beta_{fit}\rangle= 1.28\pm0.41$,
while for a double $\beta$ model, 
this value becomes 
$\langle\beta_{spec}/\beta_{fit}\rangle=0.99\pm0.35$. 
Therefore, it seems that there is no apparent $\beta$ discrepancy for clusters 
within the framework of a double $\beta$ model fit to $S_x(r)$.

\section{DISCUSSION AND CONCLUSIONS}

The conventional $\beta$ model
appears to be inaccurate, and the extrapolation of a single $\beta$ model
to large radii could be misleading. The monotonic increase of baryon 
fraction with radius and the so-called $\beta$ discrepancy,
together with the luminosity divergence as a result of the too small
$\beta_{fit}$ parameters ($\beta_{fit}\leq1/2$) for some clusters,  
may have a common origin: A single $\beta$ model fit to the entire regions of
the X-ray surface brightness profiles of clusters 
yields an underestimate of $\beta_{fit}$ parameter
at large radii. The exclusion of the central data points or the employment
of a double $\beta$ model in the fitting of the observed X-ray surface
brightness profiles  can lead to an apparent 
increase of $\beta_{fit}$ at large radii, which may allow one to
resolve all the puzzles mentioned above, although these 
two empirical methods both depend on the extent of the fitting regions.
Yet, in the conventional treatment the cooling flow regions are often 
excluded from the $\beta$ model fit, which may suffer from the uncertainty 
of different strategies of choice of the cutoff radii 
(e.g. Jones \& Forman 1984; White, Jones \& Forman 1997; Vikhlinin, Forman
\& Jones 1999). As a result, the accuracy of our determination of the total 
cluster mass from X-ray measurements may be correspondingly affected.
Employing the second $\beta$ model to quantitatively describe 
the central excess X-ray emission can significantly improve 
the goodness of the fit, which thus enables us to
derive more accurately the gas and total dynamical masses of clusters.  
In the present paper, we have extensively studied
the properties of a double $\beta$ model, and demonstrated its 
applications using two examples: A2597 and A2390.
In particular, as compared with a single $\beta$ model, 
a double $\beta$ model via the hydrostatic equilibrium 
hypothesis for intracluster gas gives rise to a larger
X-ray cluster mass at the central region, although this correction is still 
insufficient to resolve the discrepancy between
the strong lensing and X-ray determined cluster masses enclosed within the
arc radius (e.g. Wu et al. 1998).
It is hoped that the mathematical treatments of a double $\beta$ model
developed in the present paper will be useful for future work.

The $\beta$ parameter as an indicator of the dynamical properties of
clusters discussed in the present paper and  the baryon 
fraction of clusters as an indicator of cosmological matter 
composition discussed in literature (White et al. 1993; David, 
Jones \& Forman 1995; Wu \& Xue 2000 and references therein) concern mainly
the asymptotic behavior of clusters at large radii, 
while a double $\beta$ model provides a more precise way to describe 
the variation of intracluster gas at large radii. 
We have thus studied the possibility of resolving the 
well-known $\beta$ discrepancy using a double model fit 
to the X-ray surface brightness of clusters. We have found,
based a sample of 33 clusters drawn from a recent catalog of MME,
that the spectroscopic parameter $\beta_{spec}$ is on average consistent
with the observationally fitted parameter $\beta_{fit}$ from the X-ray surface
brightness profile.

Of course, the definite resolution to the $\beta$ discrepancy also needs
a better understanding of the distributions of cluster galaxies and their
velocity dispersion at large radii (Bahcall \& Lubin 1994; 
Girardi et al. 1996), for which we have assumed a simplified King model
with a constant and isotropic velocity dispersion in the present paper.  
A shallower galaxy density profile (e.g. $n_{gal}\propto r^{-2.4}$) 
can give rise to a higher value of $\beta_{fit}$. 
Alternatively, the second X-ray core radius will become larger in a double
$\beta$ model fit.
In our cluster sample, the mean core radii of the narrow and extended 
components are $\langle r_{c1}\rangle=0.14\pm0.14$ Mpc and 
$\langle r_{c1}\rangle=0.48\pm0.38$ Mpc, respectively. Since the 
extended component is involved in the evaluation of $\beta_{fit}$,
a good approximation of $\beta_{fit}$ can only be achieved at very 
large radii. So, the effect of finite observational sizes may cause
another uncertainty, and the extension of the fitting region will  
lead to an increase of $\beta_{fit}$. 
As a whole, these two effects, the shallower galaxy density profile and the
extension of the X-ray fitting region, can both result in
a high $\beta_{fit}$. Therefore, it is possible that  
the $\beta$ discrepancy still remains but  $\beta_{fit}>\beta_{spec}$.

Finally, within the framework of energy equipartition 
the ratio of kinetic energy
in the galaxies to that in the gas should be close to unity. Although our
estimate of $\langle\beta_{spec}\rangle=0.78\pm0.37$ is still
consistent with such a prediction within the error bars, this value, 
together with the consistent $\beta_{fit}$ parameter 
$\langle\beta_{fit}\rangle=0.83\pm0.33$, does not exclude the possibility
that  the actual value of $\langle\beta_{spec}\rangle$ is smaller than 1,
namely, $\langle\sigma^2\rangle<\langle kT\rangle/\mu m_p$.  
Here, we will not intend to explore in detail the physical mechanisms 
for the scenario. However, we would like to  
point out that the dynamical friction would resist the motion of galaxies
in clusters, leading to a velocity dispersion smaller than that 
expected from the purely dissipationless hypothesis 
(e.g. Carlberg \& Dubinski 1991; Metzler \& Evrard 1994). This may
account for our result -- why galaxies appear to be cooler than 
intracluster gas.

%\vskip 1cm

\section*{Acknowledgments}

We thank  Haiguang Xu for useful discussion and help with 
the understanding of the ROSAT PSPC 
data reduction, and an anonymous referee for many valuable suggestions
and comments. This work was supported by 
the National Science Foundation of China, under Grant No. 1972531.

%\newpage

\newpage

\begin{figure*}
\centerline{\hspace{5cm}\psfig{figure=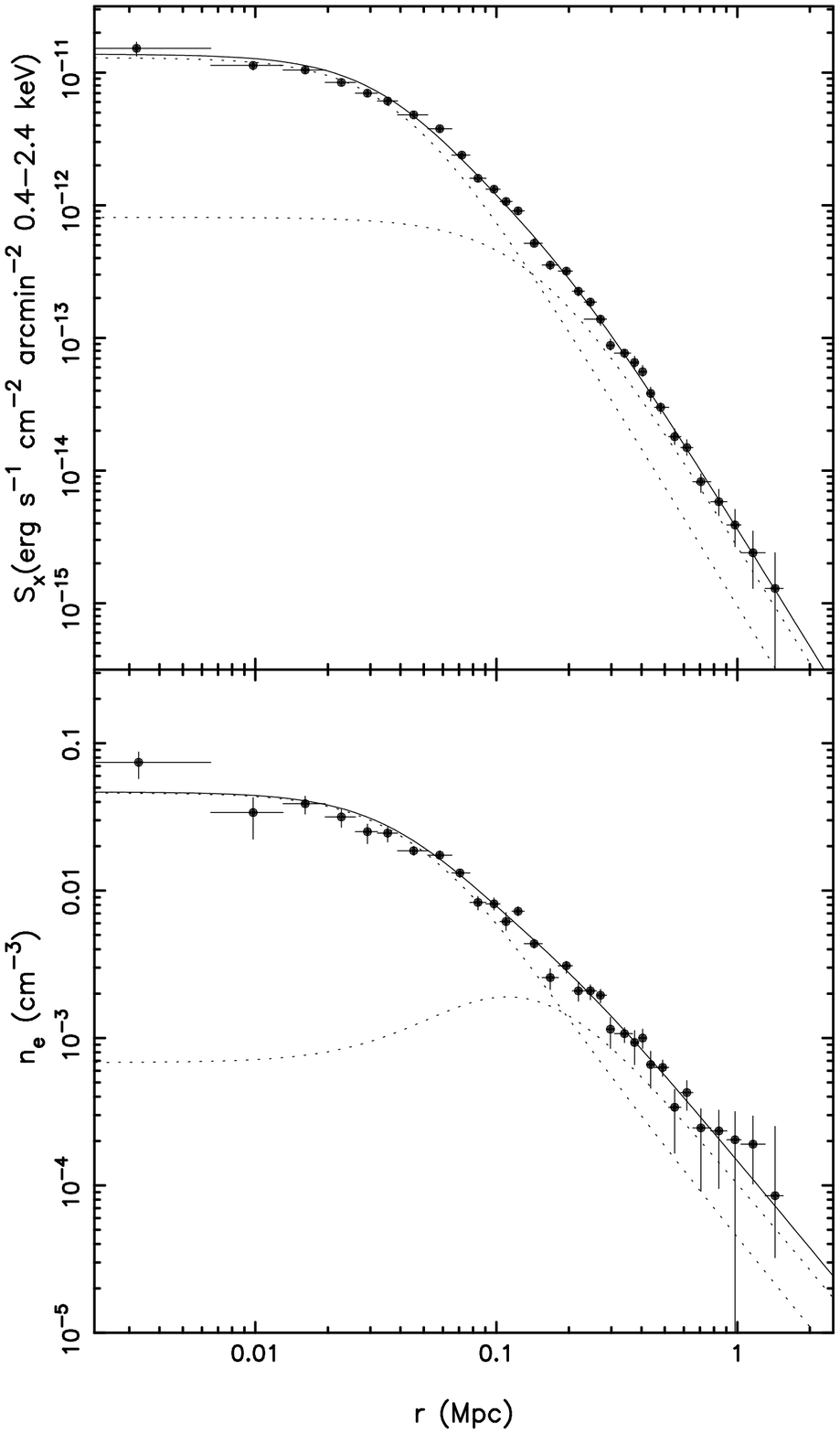,width=1.0\textwidth,angle=0}}
\caption{
Upper panel: The merged ROSAT HRI and PSPC surface brightness
profile for A2579 (Sarazin \& McNamara 1997), along with 
the best-fit double $\beta$ model (solid line). The effect of the 
PSF of the ROSAT HRI has been corrected.   
Lower panel: A comparison of the electron number density derived from
the deprojection technique (Sarazin \& McNamara 1997) and that from
the best-fit double $\beta$ model (solid line). 
The dotted lines represent the two components of the double $\beta$ model.
}
\end{figure*}

\begin{figure*}
\centerline{\hspace{3cm}\psfig{figure=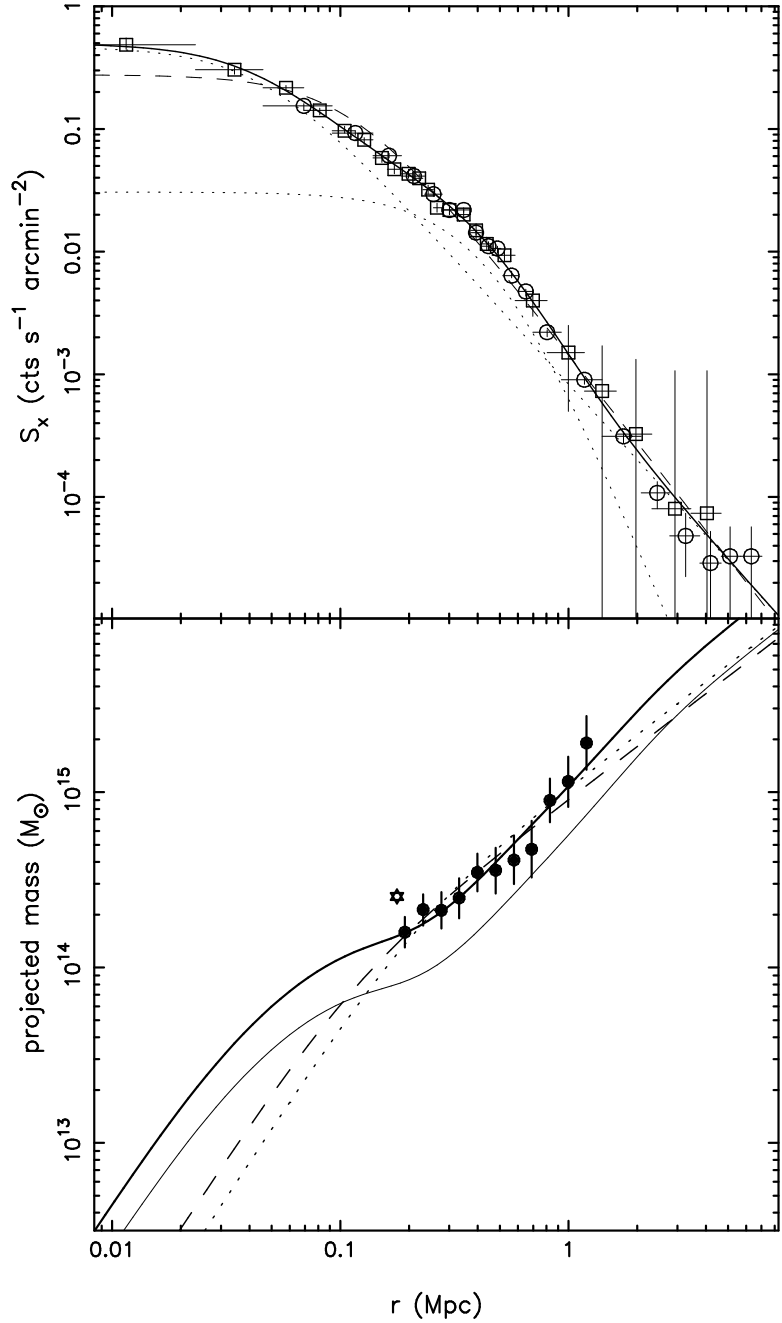,width=1.\textwidth,angle=0}}
\caption{
Upper panel: The superimposed ROSAT HRI (squares) and PSPC (circles)
surface brightness profiles for A2390 (B\"ohringer et al. 1998).
Also plotted are the best-fit double $\beta$ model (solid line) and 
the best-fit single $\beta$ model (dashed line) for the entire data points.
The dotted lines represent the two components of the double $\beta$ model.
Lower panel: A comparison of the projected cluster masses 
revealed by weak gravitational lensing (filled circles) 
(Squires et al. 1996) and by the X-ray measurements via 
hydrostatic equilibrium hypothesis: thin solid line -- 
the best-fit double $\beta$ model with $T_1=2$ keV and $T_2=11.5$ keV;
thick solid line -- 
the best-fit double $\beta$ model with $T_1=4.32$ keV and $T_2=20.0$ keV;
dashed line -- the best-fit 
single $\beta$ model for the entire data points; dotted line --
the single $\beta$ model fit by exclusion of the central emission excess.
The strong lensing derived cluster mass is also shown by asterisk
(Wu et al. 1998).
}
\end{figure*}

\begin{figure*}
\centerline{\hspace{5cm}\psfig{figure=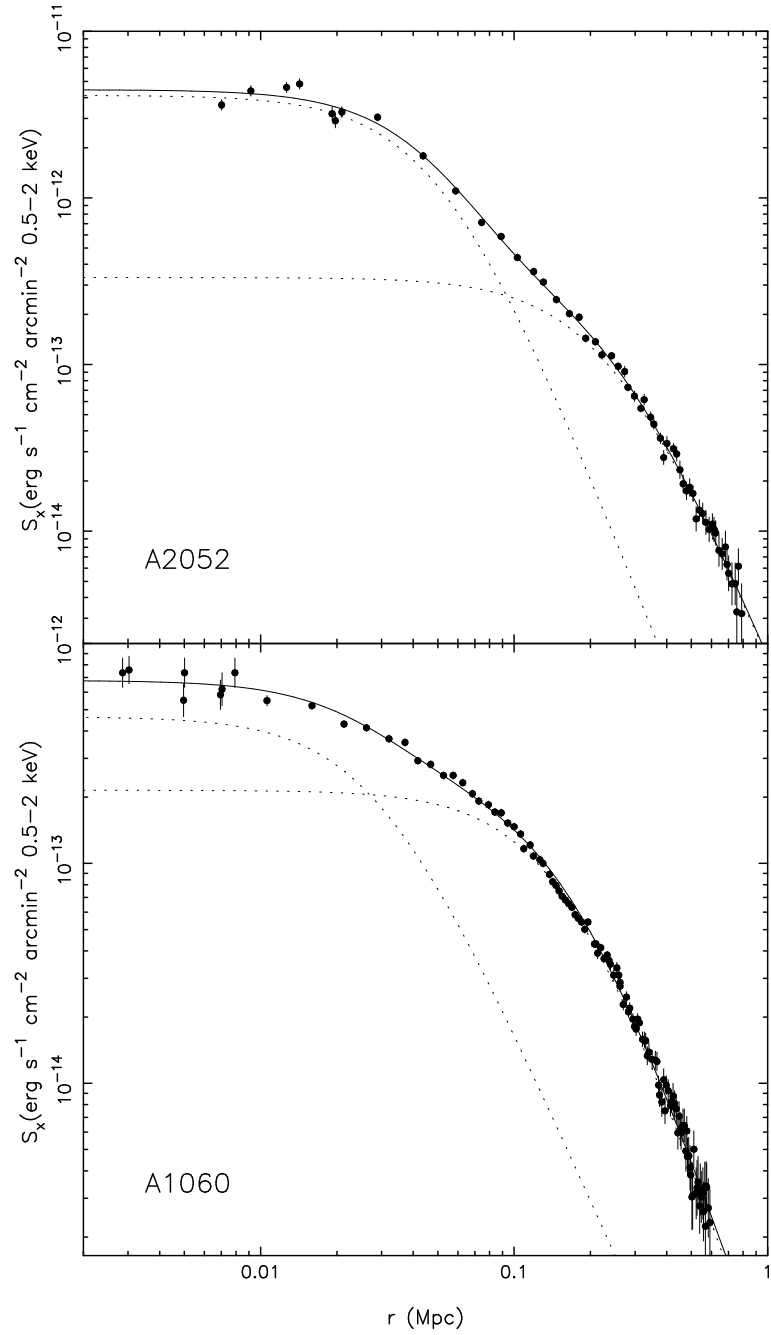,width=1.0\textwidth,angle=0}}
\caption{Two typical examples of the ROSAT PSPC observed and 
the  double $\beta$ model fitted surface brightness profiles, A2052
and A1060. 
The dotted lines represent the two components of the double $\beta$ model.  
}
\end{figure*}

\begin{figure*}
\centerline{\hspace{3cm}\psfig{figure=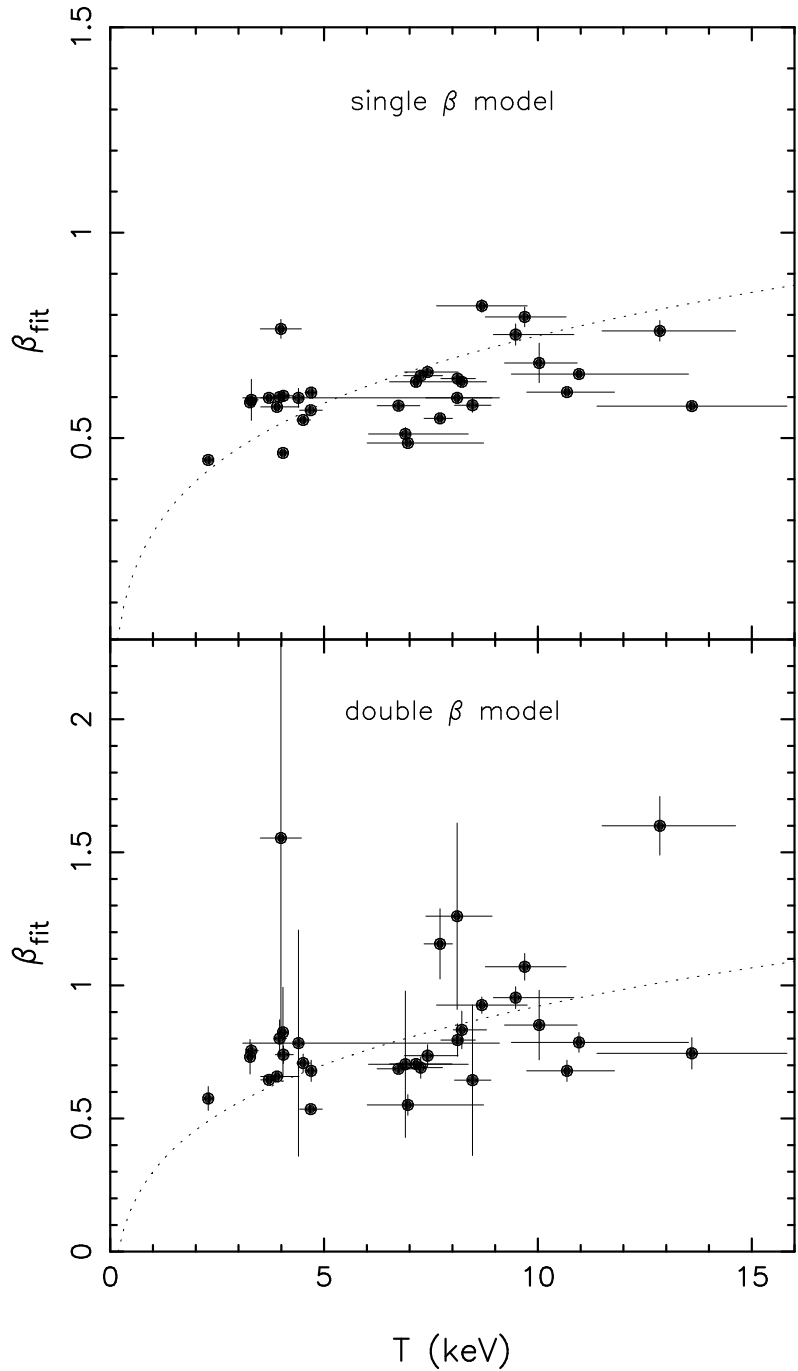,width=1.\textwidth,angle=0}}
\caption{The $\beta$ parameters obtained from a single (upper panel) and
double (lower panel) $\beta$ model fit plotted against the X-ray temperature.
The dotted lines are the best-fit correlations represented by the 
form $\beta=T^a+b$, in which $a= 0.17\pm0.03$ and $0.22\pm0.03$ for
a single and double $\beta$ model, respectively.
Cygnus-A is not shown in the plot due to its high temperature (39.40 keV). }
\end{figure*}

\begin{figure*}
\centerline{\hspace{3cm}\psfig{figure=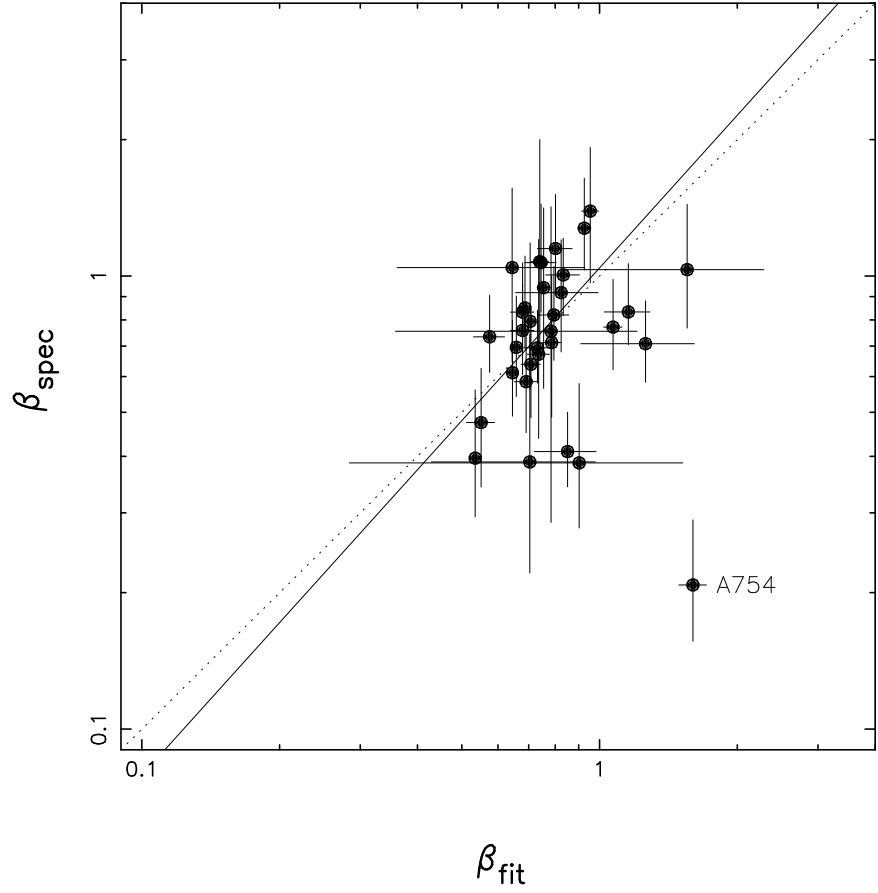,width=1.3\textwidth,angle=270}}
\caption{The spectroscopic parameter $\beta_{spec}$ is plotted against
the observationally fitted parameter $\beta_{fit}$ from $S_x(r)$ for our 
sample of 33 clusters. The solid line is the best $\chi^2$-fit 
power-law relation to the data set with exclusion of A754: 
$\beta_{spec}\propto \beta_{fit}^{1.12}$, while the dotted line
denotes $\beta_{spec}= \beta_{fit}$.}
\end{figure*}

\begin{figure*}
\centerline{\hspace{3cm}\psfig{figure=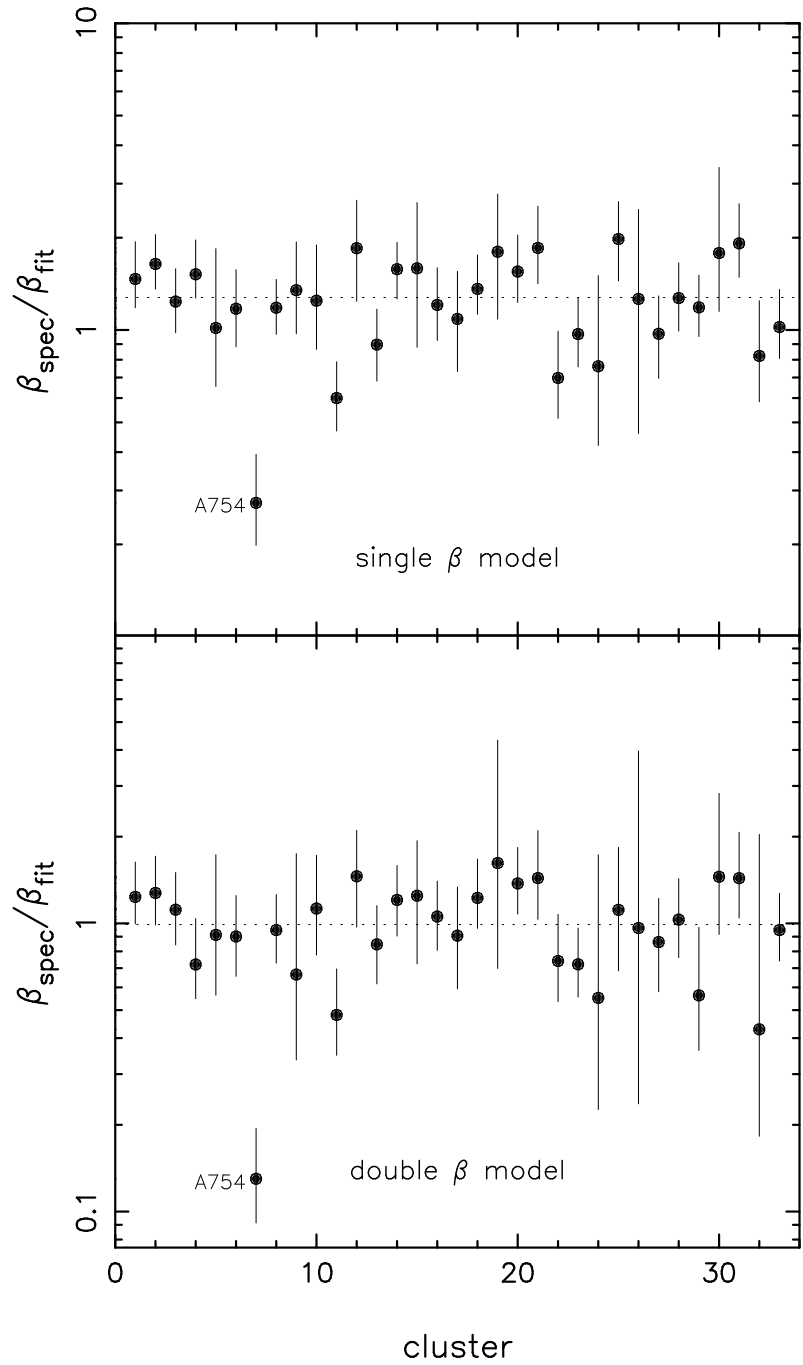,width=1.\textwidth,angle=0}}
\caption{The ratio of $\beta_{spec}$ to $\beta_{fit}$ for an ensemble 
of 33 clusters. Upper and lower panels correspond to a single and double
$\beta$ model fit, respectively. The dotted lines denote the mean ratios.}
\end{figure*}

\end{document}